\documentclass[11pt,preprint]{aastex}
\usepackage{graphicx}

\begin{document}

\title{Hot Molecular Circumstellar Disk around Massive Protostar Orion Source~I}

\author{Tomoya HIROTA\altaffilmark{1,2}, 
Mi Kyoung KIM\altaffilmark{3}, 
Yasutaka KURONO\altaffilmark{4,5}, 
Mareki HONMA\altaffilmark{1,2}}
\email{tomoya.hirota@nao.ac.jp}
\altaffiltext{1}{National Astronomical Observatory of Japan, Mitaka, Tokyo 181-8588, Japan}
\altaffiltext{2}{Department of Astronomical Sciences, Graduate University for Advanced Studies, Mitaka, Tokyo 181-8588, Japan}
\altaffiltext{3}{Korea Astronomy and Space Science Institute, Hwaam-dong 61-1, Yuseong-gu, Daejeon, 305-348, Republic of Korea}
\altaffiltext{4}{Chile Observatory, National Astronomical Observatory of Japan, Osawa 2-21-1, Mitaka, Tokyo 181-8588, Japan}
\altaffiltext{5}{Joint ALMA Observatory, Alonso de Cordova 3107 Vitacura, Santiago 763-0355, Chile}

\begin{abstract}
We report new Atacama Large Millimeter/Submillimeter Array (ALMA) observations of a circumstellar disk around Source~I in Orion~KL, an archetype of massive protostar candidate. 
We detected two ortho-H$_{2}$O lines at 321~GHz ($10_{2,9}$-$9_{3,6}$) and 336~GHz ($\nu_{2}=1, 5_{2,3}$-$6_{1,6}$) for the first time in Source~I. 
The latter one is in vibrationally excited state at the lower state energy of 2939~K, suggesting an evidence of hot molecular gas close to Source~I. 
The integrated intensity map of the 321~GHz line is elongated along the bipolar outflow while the 336~GHz line map is unresolved with a beam size of 0.4\arcsec. 
Both of these maps show velocity gradient perpendicular to the bipolar outflow. 
The velocity centroid map of the 321~GHz line implies spatial and velocity structure similar to that of vibrationally-excited SiO masers tracing the root of the outflow emanating from the disk surface. 
In contrast, the 336~GHz line is most likely emitted from the disk midplane with a diameter of 0.2\arcsec (84~AU) as traced by a radio continuum emission and a dark lane devoid of the vibrationally-excited SiO maser emission. 
The observed velocity gradient and the spectral profile of the 336~GHz H$_{2}$O line can be reconciled with a model of an edge-on ring-like structure with an enclosed mass of $>$7$M_{\odot}$ and an excitation temperature of $>$3000~K. 
The present results provide a further evidence of hot and neutral circumstellar disk rotating around Source~I with a diameter of $\sim$100~AU scale. 
\end{abstract}


\section{Introduction}

While it is well understood how a low-mass Solar-type star forms, the formation process of massive stars remains yet to be addressed.
Recent observations of massive young stars suggested existence of circumstellar disks around them \citep{Patel2005,Jiang2005,Kraus2010}, which favors a formation via disk accretion rather than other processes such as stellar collisions \citep{Bally2005}. 
However, physical conditions of accretion disks, especially at a scale of $\sim$100~AU, are still debatable mainly due to limited spatial resolution. 

Because of its proximity to the Sun at a distance of 420~pc \citep{Menten2007, Kim2008}, Orion~KL (Kleinman-Low object) has been recognized as one of the best laboratories to study massive star-formation processes. 
Among a number of young stars, a radio source called Source~I is prominent in this region and is thought to have a high luminosity of $>$10$^{4}L_{\odot}$ \citep{Menten1995}. 
It is the driving source of a bipolar outflow along the northeast-southwest direction with a scale of 1000~AU \citep{Wright1995,Zapata2012,Niederhofer2012,Greenhill2013}. 
At the launching point of the bipolar outflow, there is a cluster of vibrationally-excited SiO masers tracing an outflow arising from the surface of the disk with a diameter of $\sim$100~AU, revealed by VLBI observations with resolutions of 0.8~milliarcseconds (mas) \citep{Kim2008} and 0.5~mas \citep{Matthews2010}. 
A compact radio continuum source is associated at the center of the vibrationally-excited SiO masers which is interpreted as an edge-on disk \citep{Reid2007,Goddi2011,Plambeck2013}. 

Nevertheless, the nature of Source~I and associated disk/outflow system are still far from a complete understanding. 
One of the long-standing issues is an origin of the radio continuum emission. 
While the 43~GHz continuum emission is elongated perpendicular to the northeast-southwest outflow \citep{Wright1995,Zapata2012,Niederhofer2012,Greenhill2013}, it seems consistent with the base of a larger scale high-velocity outflow traced by near-infrared shocked H$_{2}$ emission \citep{Allen1993,Bally2011}. 
For this reason, it has sometimes been interpreted as a radio jet toward northwest-southeast directions \citep{Testi2010,Chatterjee2012}. 

Furthermore, there are continued discussions related to the mass of Source~I. 
A rotation curve of the vibrationally-excited 43~GHz SiO masers indicates an enclosed mass of $>$(7-8)$M_{\odot}$ \citep{Kim2008,Matthews2010}. 
Near-infrared reflected spectrum can be explained by an emission from a circumstellar disk around a 10$M_{\odot}$ protostar \citep{Testi2010}. 
On the other hand, it has been claimed that Source~I would consist of a massive binary system with a total mass of 20$M_{\odot}$, given the dynamics of this region including Source~I and BN (Becklin-Neugebauer) object \citep{Gomez2008,Bally2011,Goddi2011}, although these dynamical scenarios are still controversial \citep{Chatterjee2012}. 

Physical properties of the associated disk are also under debate \citep{Reid2007,Testi2010,Bally2011,Goddi2011,Okumura2011,Plambeck2013,Sitarski2013}, which are crucial to understand massive star-formation mechanism. 
Further observations of Source~I are essential to understand not only its origin but also formation mechanisms of massive stars. 
For this purpose, we here present observational results of submillimeter H$_{2}$O lines detected in  Source~I with newly constructed Atacama Large Millimeter/Submillimeter Array (ALMA). 

\section{Observations}

Observations were carried out with ALMA on July 16, August 25, and October 21, 2012. 
The target source was Orion~KL and the tracking center position was taken to be the bursting 22~GHz H$_{2}$O maser, RA(J2000)=05h35m14.125s, Decl(J2000)=-05d22\arcmin36.486\arcsec \citep{Hirota2011}, which is 7\arcsec \ southwest of Source~I. 
The on-source integration time was about 100~seconds for each session. 
The array was consisted of 21-28 antennas with a diameter of 12~m each in the extended configuration with the maximum baseline length of 400~m. 
A primary flux calibrator, band-pass calibrator, and secondary gain calibrator were Callisto, J053851-440507/J0423-013, and J0607-085, respectively. 

Two ortho-H$_{2}$O lines at 321.225656~GHz ($10_{2,9}$-$9_{3,6}$) and 336.227931~GHz ($\nu_{2}=1, 5_2,3$-$6_{1,6}$) were observed simultaneously. 
The 336~GHz H$_{2}$O line lies in the excited state in the bending mode ($\nu_{2}$=1). 
The lower state energy levels are 1846~K and 2939~K for the 321~GHz and 336~GHz H$_{2}$O lines, respectively \citep{Chen2000}. 
The ALMA correlator provided four spectral windows with a bandwidth of 468.750~MHz for each. 
The channel spacing of spectrometers was 122~kHz. 
The system noise temperature ranged from 100-200~K, depending on the observing frequency and weather conditions. 

\section{Data reduction}

Synthesis imaging and self-calibration were done with the CASA software package. 
We set a velocity resolution of 0.125~km~s$^{-1}$ (corresponds to 130-140~kHz resolution) in the synthesized imaging of the H$_{2}$O lines. 
First, both phase and amplitude self-calibrations were done with the continuum emission of Orion~KL by integrating over line-free channels. 
Peak intensity and rms noise level of the continuum emission are 698~mJy~beam$^{-1}$ and 9~mJy~beam$^{-1}$, respectively. 

\begin{figure}
\begin{center}
\includegraphics[width=9cm]{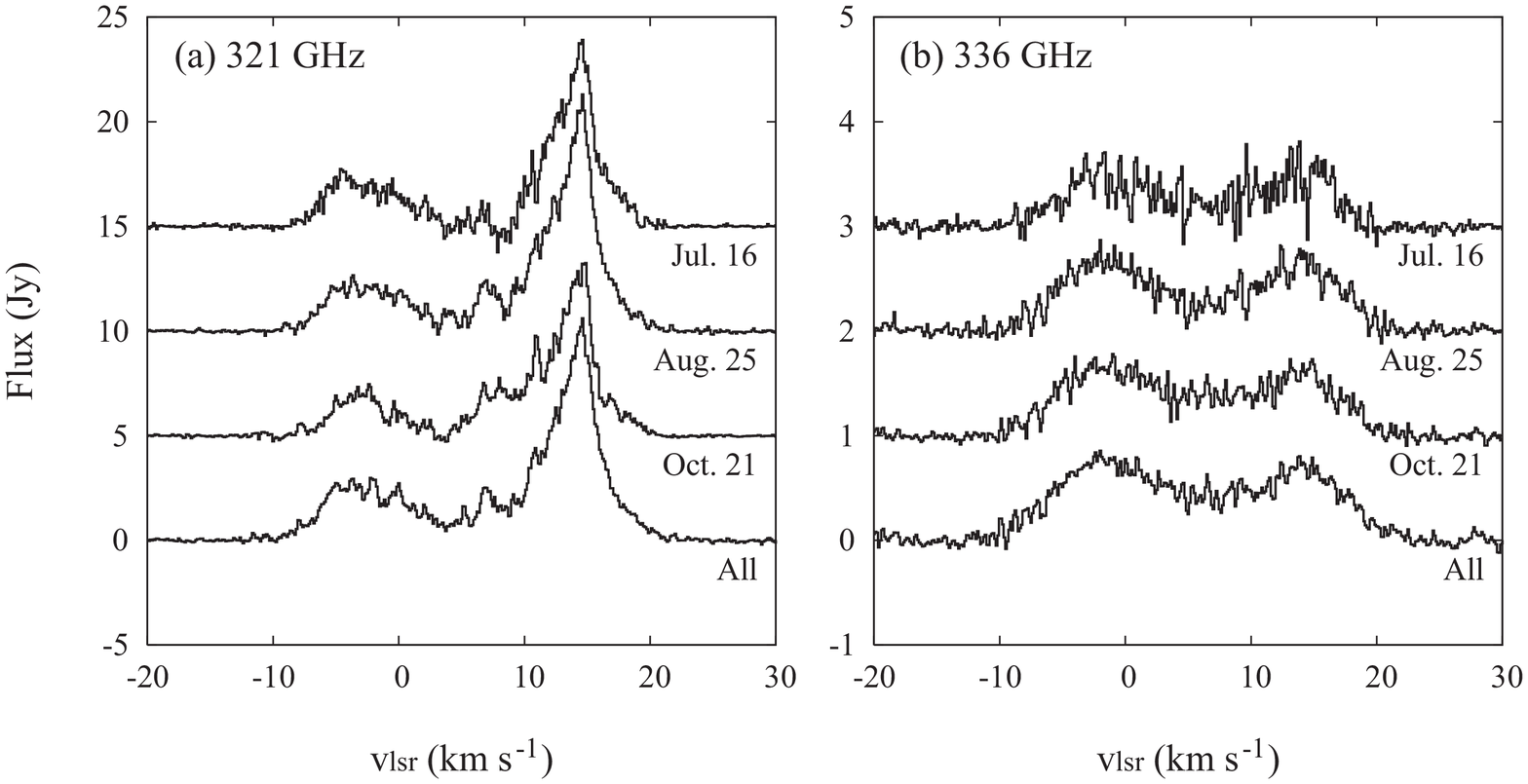}
\caption{
Observed spectra of the (a) 321~GHz and (b) 336~GHz H$_{2}$O lines for those at each epoch and integrated over all of the three epochs. 
}
\label{fig-sp}
\end{center}
\end{figure}

The solutions of self-calibration were applied to all the spectral channels including the target H$_{2}$O lines. 
We compared the continuum emission and selected spectral lines of methyl formate, HCOOCH$_{3}$ to check the stability of the observed flux scale. 
As a result, we found a possible variation in continuum and line intensities between different observing sessions. 
It is most likely due to different array configurations which could result in different degrees of missing flux for extended emission features. 
On the other hand, we could not find significant flux variation of the 321~GHz and 336~GHz H$_{2}$O lines as shown in Figure \ref{fig-sp}. 
This suggests that the H$_{2}$O lines are emitted from a compact source and have no intrinsic time variation unlike circumstellar 321~GHz H$_{2}$O masers around late-type stars \citep{Yates1996}. 
Considering these results, we estimated an accuracy of flux measurements to be $\sim$20\%. 

In order to obtain higher spatial resolution and to exclude contribution from extended emission component, we produced synthesized images of the 321~GHz and 336~GHz H$_{2}$O lines as well as the continuum emission by using the uniform-weighted visibility data with a UV distance longer than 200~k$\lambda$. 
The resultant beam size is 0.40\arcsec$\times$0.34\arcsec \ with a position angle of 60~degrees. 
The image rms noise level was $\sim$30~mJy~beam$^{-1}$ at each channel. 
Because these maps are free from the sidelobes caused by the strong extended emission, the noise levels are improved compared with those of full UV sampling images. 

\section{Results}

\begin{figure}
\begin{center}
\includegraphics[width=9cm]{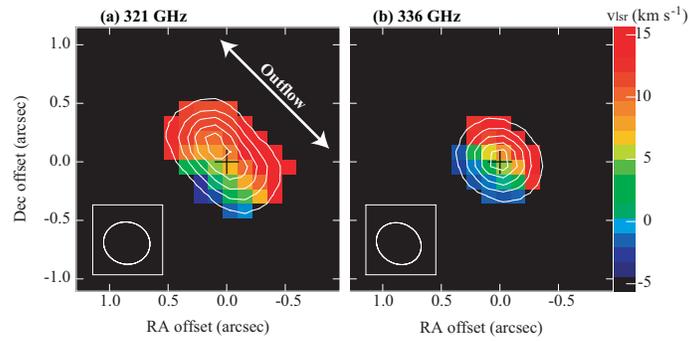}
\caption{
(a) Integrated intensity (contours) and radial velocity (color) of the 321~GHz H$_{2}$O line. 
The contour levels are 10,30,50,70, and 90\% of the peak intensity of 48.1~Jy~beam$^{-1}$. 
The (0,0) position is taken from the 321/336~GHz continuum peak, RA=05h35m14.512s and Decl=-05$^{\circ}$22\arcmin30.57\arcsec (J2000) as indicated by a black cross. 
Directions of the bipolar outflow is indicated by a white arrow \citep{Wright1995,Zapata2012,Niederhofer2012,Greenhill2013}. 
A synthesized beam size is shown in the bottom-left corner. 
(b) Same as (a) but for the 336~GHz H$_{2}$O line. 
The peak intensity of the 336~GHz H$_{2}$O line is 14.4~Jy~beam$^{-1}$. 
}
\label{fig-map}
\end{center}
\end{figure}

We detected two H$_{2}$O lines at 321~GHz and 336~GHz toward Source~I (Figure \ref{fig-sp}). 
Although the 321~GHz line has been detected in star-forming regions \citep{Menten1991,Patel2007}, the vibrationally-excited 336~GHz line has been detected only in a red supergiant such as VY~CMa \citep{Menten2006}. 
Thus, this is the first detection of the 336~GHz H$_{2}$O line in star-forming regions. 
Along with another vibrationally-excited H$_{2}$O line at 232~GHz newly detected with ALMA \citep{Hirota2012}, these transitions will be unique tracers of massive young stars. 
Both of the 321~GHz and 336~GHz emissions are concentrated in a close vicinity of Source~I. 
Because of their high excitation energy levels, these lines clearly suggest a hot molecular gas associated with Source~I. 
It is a striking difference from the lower-excitation 22~GHz H$_{2}$O masers (lower state energy level of 642~K) spread over the bipolar outflow with a much larger scale of 1000~AU\citep{Greenhill2013}. 
The spatial structure of the integrated intensity map of the 321~GHz H$_{2}$O line shows an elongation along the northeast-southwest direction, which is indicative of the bipolar outflow \citep{Wright1995,Zapata2012,Niederhofer2012,Greenhill2013}, 
whereas the 336~GHz map is more compact than the 321~GHz map and its extent is comparable to the present resolution (Figure \ref{fig-map}). 

To evaluate their structures, we made velocity centroid maps (peak positions of velocity channel maps) of both lines by performing two dimensional Gaussian fitting to the synthesized images. 
The positional uncertainty is proportional to the synthesized beam size and is inversely proportional to the signal-to-noise ratio of the images. 
At the peak velocity channel of the 336~GHz H$_{2}$O line, the signal-to-noise ratio is $\sim$30, and hence, the positional accuracy is estimated to be as high as 0.01\arcsec. 
Formal errors in the Gaussian fitting, 0.005-0.02\arcsec (1$\sigma$), are almost consistent with this expectation. 

\begin{figure*}
\begin{center}
\includegraphics[width=17cm]{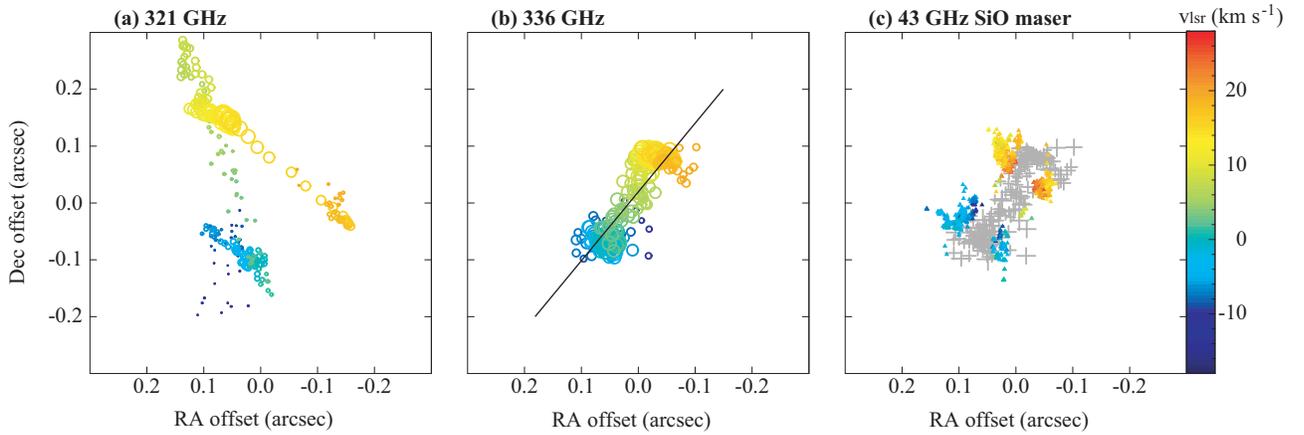}
\caption{
(a) Velocity centroid map of the 321~GHz H$_{2}$O line. 
The color represents the radial velocity and the size of the circle is proportional to its intensity. 
The typical values of positional errors in the Gaussian fitting is 0.005-0.02\arcsec (1$\sigma$). 
(b) Same as (a) but for the 336~GHz H$_{2}$O line. 
A solid line with a position angle of 140$\pm1$~degrees represents the midplane of the disk derived from peak positions of these channel maps. 
(c) Same as (a) but for the 43~GHz vibrationally-excited SiO maser lines observed with VERA (VLBI Exploration of Radio Astrometry) with 0.8~mas$\times$0.5~mas resolution \citep{Kim2008}. 
Gray crosses represent peak positions of the 336~GHz channel maps (panel (b)) with error bars of two-dimensional Gaussian fitting of each channel map (1$\sigma$). 
}
\label{fig-centroid}
\end{center}
\end{figure*}

The 321~GHz map shows an inverted Z-shaped structure with sharp ridges at northwestern and southeastern sides connected by linearly aligned features (Figure \ref{fig-centroid}(a)). 
This appears quite similar to previous interferometer maps of vibrationally-excited ($v$=1, 2) $^{28}$SiO masers, as well as ground state ($v$=0) $^{29}$SiO and $^{30}$SiO masers, at 43~GHz and 86~GHz \citep{Menten1995,Wright1995,Baudry1998,Goddi2009}. 
Thus, the vibrationally-excited SiO masers and the 321~GHz H$_{2}$O line trace the same dynamical structure excited at the base of the northeast-southwest outflows \citep{Kim2008,Matthews2010}. 
In fact, the 321~GHz H$_{2}$O line is detected in a bipolar outflow associated with another nearby massive star-forming region Cepheus~A \citep{Patel2007}. 
The resemblance of 321~GHz H$_{2}$O emission and previous vibrationally-excited SiO maser observations ensures that the distribution of the H$_{2}$O emission traced by ALMA is reliable and can be compared with higher angular resolution maps of VLBI observations. 

On the other hand, the 336~GHz H$_{2}$O line map perfectly fits the dark lane devoid of the vibrationally-excited SiO masers in an X-shaped distribution obtained with higher resolution VLBI at 43~GHz (Figure \ref{fig-centroid}(b)(c)) \citep{Kim2008,Matthews2010}. 
The distribution of the 336~GHz H$_{2}$O channel maps, $\sim$0.2\arcsec$\times$0.1\arcsec (84AU$\times$42AU), is comparable with the size of the 43~GHz continuum emission \citep{Reid2007,Goddi2011}. 
It is most likely that the 336~GHz H$_{2}$O line is emitted from a midplane of the disk as traced by the 43~GHz continuum emission. 

The most important finding is a clear velocity gradient along the northwest-southeast direction (Figure \ref{fig-pv}(a)). 
The gradient is analogous to that of the vibrationally-excited SiO masers \citep{Kim2008,Matthews2010}, and is perpendicular to the northeast-southwest outflow \citep{Wright1995,Zapata2012,Niederhofer2012,Greenhill2013}. 
The quasi-linear velocity gradient with a lack of the highest velocity components close to the central position suggests that the 336~GHz H$_{2}$O line is emitted from a rotating ring-like structure or limb of the disk in an edge-on view.
If we simply assume a linear velocity gradient of the rotating edge-on ring with a radius of 42~AU, the enclosed mass is estimated to be 5$M_{\odot}$. 
Although this value is smaller than the previous estimate of $>$(7-8)$M_{\odot}$ \citep{Kim2008,Matthews2010}, 
a slightly larger radius of 47~AU yields a consistent value of 7$M_{\odot}$. 
These results are much lower than another estimate, 20$M_{\odot}$, according to the conservation of momentum of Source~I and BN object estimated from their proper motions \citep{Gomez2008, Goddi2011}. 
Because the mass derived from the velocity gradient, (5-7)$M_{\odot}$, is regarded as a rotationally supported part of the system, it could be a lower limit of the total mass of Source~I. 
If this is the case, it implies that non-gravitational forces such as magnetic field and/or radiation pressure would efficiently support the Source~I system \citep{Bally2011,Goddi2011}. 

One might claim that the H$_{2}$O lines could trace a bipolar outflow along the northwest-southeast direction \citep{Greenhill1998,Testi2010,Chatterjee2012}. 
However, if the H$_{2}$O lines and radio continuum emission trace the same volume of this outflow, the expansion velocity would be $\sim$20~km~s$^{-1}$ assuming a moderate inclination of the outflow axis of 45~degrees. 
It corresponds to a proper motion of 0.01~arcsec~yr$^{-1}$, which is inconsistent with a stable structure of the 43~GHz continuum emission \citep{Goddi2011}. 
A radial expansion or infall are also unlikely because of a lack of the highest radial velocity components toward the central position in the observed position-velocity (PV) diagram (Figure \ref{fig-pv}(a)) and a broad wing components in the 336~GHz H$_{2}$O line spectrum (Figure \ref{fig-pv}(b)). 
Thus, 336~GHz H$_{2}$O line map provides a further evidence of a rotating circumstellar disk around Source~I. 

\section{Discussion}

In order to reconstruct observed PV diagram and spectral profile of the 336~GHz H$_{2}$O line, we here employ a model of an edge-on rotating disk assuming a ring-like structure with uniform temperature/density for simplicity. 
Because of a limited spatial resolution (0.4\arcsec=168~AU), it is hard to solve a degeneracy of disk parameters, such as mass, size, and temperature, based on our present observations with a single transition at 336~GHz. 
Thus, what we have done is a proof-of-concept model calculation to demonstrate that the observed PV diagram and spectral profile can be indeed explained with reasonable physical and dynamical parameters consistent with previous observations. 

\begin{figure}
\begin{center}
\includegraphics[width=17cm]{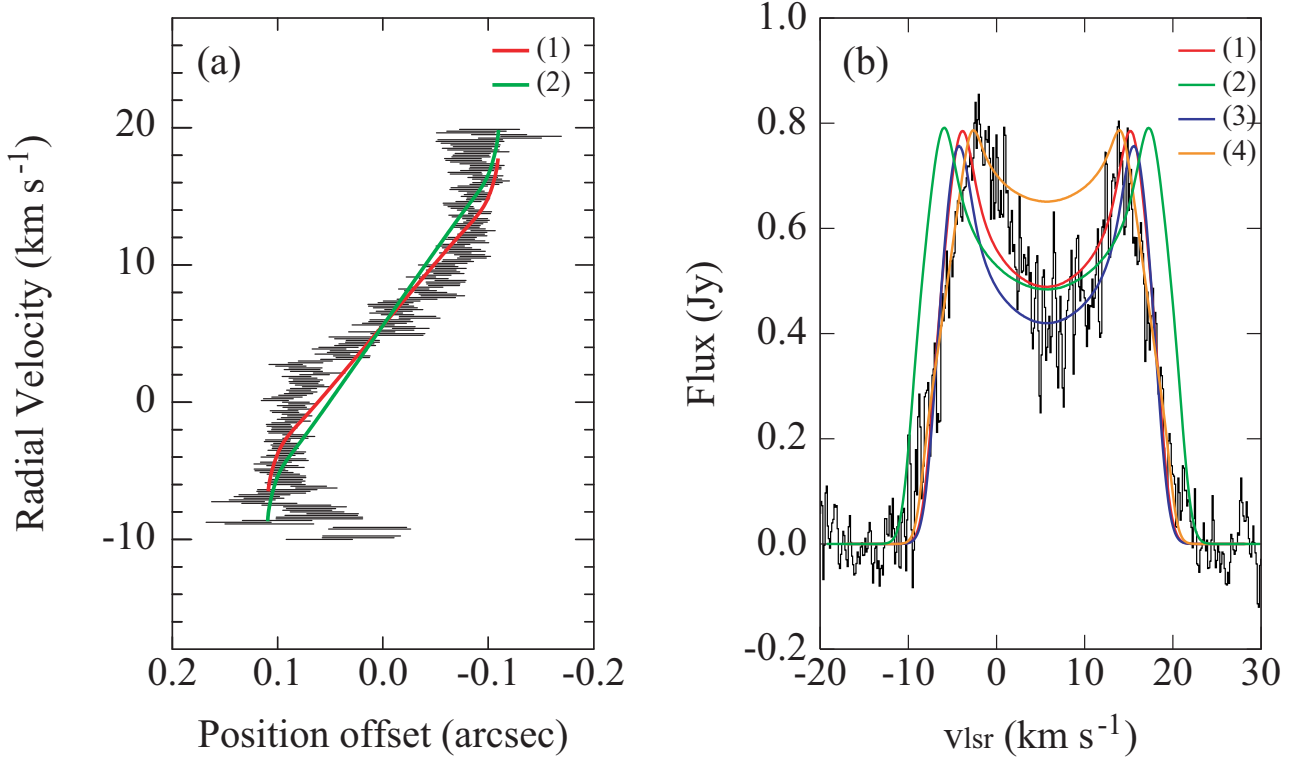}
\caption{
(a) Position-velocity (PV) diagram of the 336~GHz H$_{2}$O line along the disk midplane indicated by the solid line in Figure \ref{fig-centroid} (b). 
Observed PV diagram is shown by horizontal bars indicating formal errors in positions (1$\sigma$) of two-dimensional Gaussian fitting of each channel map. 
Models are also shown with parameters of (1) $r_{\mbox{in}}$=45~AU, $r_{\mbox{out}}$=50~AU, $T_{\mbox{ex}}$=3000~K, and $M_{\mbox{rot}}$=7$M_{\odot}$, (2) $r_{\mbox{in}}$=45~AU, $r_{\mbox{out}}$=50~AU, $T_{\mbox{ex}}$=3000~K, and $M_{\mbox{rot}}$=10$M_{\odot}$. 
(b) Observed and model spectra of the 336~GHz H$_{2}$O line. 
Parameters for model spectra (1) and (2) are the same as employed in panel (a). 
We added models with (3) same as (1) but with  $T_{\mbox{ex}}$=4500~K and (4) same as (1) but with  $T_{\mbox{ex}}$=1500~K. 
}
\label{fig-pv}
\end{center}
\end{figure}

First, we calculated models for several parameter set of inner and outer radii of the ring, $r_{\mbox{in}}$=5, 15, 25, 35, 45, 55~AU and $r_{\mbox{out}}$=40, 50, and 60~AU with a constraint of $r_{\mbox{out}}>r_{\mbox{in}}$, for a fixed enclosed mass $M_{\mbox{rot}}$=7$M_{\odot}$ (rotationally supported mass) consistent with previously estimated lower limits \citep{Kim2008,Matthews2010}. 
As a result, the model with $r_{\mbox{in}}$=45~AU and $r_{\mbox{out}}$=50~AU is found to well reproduce the observed PV diagram and spectral profile (i.e. model (1) in Figure \ref{fig-pv}). 
If we assume smaller inner radii, models show steeper velocity gradients and higher velocity components close to the central position. 
Furthermore, smaller inner radii yield broader linewidths which result in larger discrepancy between observed and model profile. 
Thus, our results prefer the ring-like structure rather than a disk without a central hole. 

Next, we verify models with different enclosed masses. 
When we employ a larger mass of $M_{\mbox{rot}}$=10$M_{\odot}$, the velocity gradient becomes steeper and the spectral profile becomes broader similar to the model with smaller ring size (model (2) in Figures \ref{fig-pv}). 
When we fit the model spectrum with $M_{\mbox{rot}}$=10$M_{\odot}$ to the observed line profile, we should have assumed the larger disk size of $r_{\mbox{out}}>$70~AU. 
However, widths of position offsets of such model PV diagrams became significantly larger than the observed results. 
Therefore, our results prefer lower value of $\sim$7$M_{\odot}$ consistent with the previous VLBI results \citep{Kim2008,Matthews2010}. 

Finally, we estimate the excitation temperature of the 336~GHz H$_{2}$O line, $T_{\mbox{ex}}$, by fitting the observed line profile, assuming $M_{\mbox{rot}}$=7$M_{\odot}$, $r_{\mbox{in}}$=45~AU and $r_{\mbox{out}}$=50~AU, respectively. 
We find that the observed spectral profile can be reconciled with an excitation temperature of $>$3000~K and a uniform H$_{2}$O density of 5$\times$10$^{5}$~cm$^{-3}$ (e.g. models (1) and (3) in Figure \ref{fig-pv}(b)). 
When we assume lower excitation temperatures, the dip of the double-peaked profile becomes shallower due to the increase of optical depth (model (4) in Figure \ref{fig-pv}(b)). 
Note that these values could not be constrained due to a lack of multi-transition data for an excitation analysis such as rotation diagram method. 

Although we cannot rule out a possibility of excitation of the 336~GHz H$_{2}$O line via maser action, it is proposed to be thermalized via collisional and/or radiative excitation in a red supergiant VY~CMa \citep{Menten2006}, which is consistent with a theoretical prediction \citep{Alcolea1993}. 
Under this condition, its excitation temperature directly reflects a high kinetic temperature. 
Furthermore, because existence of H$_{2}$O suggests an absence of fully ionized gas, the temperature would be lower than $\sim$4500~K as estimated from radio and infrared observations \citep{Reid2007,Testi2010,Plambeck2013}. 
The temperature range of 3000-4500~K is significantly higher than that expected for a radiative equilibrium with the expected luminosity of Source~I, 10$^{4}$-10$^{5}L_{\odot}$ ($\sim$1000~K). 
Along with the ring-like structure, it may imply a heating mechanism via accretion shock in the disk midplane \citep{Reid2007,Testi2010,Plambeck2013}. 

Alternatively, the ring-like structure may trace the edge of the disk because of a high opacity of the disk even at 336~GHz. 
In fact, the continuum emission of Source~I, which could be interpreted as an H$^{-}$ free-free emission \citep{Reid2007, Plambeck2013}, shows a spectral energy distribution with a power-law index close to 2 (Hirota et al. in preparation). 
Higher frequency observations of continuum and H$_{2}$O lines will discriminate these hypothesis. 

\section{Summary}

We detected a hot neutral circumstellar disk rotating around Source~I with a scale of 100~AU in diameter. 
We derived the rotationally supported enclosed mass of 7$M_{\odot}$, which could be a lower limit of the total mass of the Source~I system. 
Because the fully ionized H{\sc{ii}} region is not significantly evolved, Source~I would be in a very early phase of massive protostar being formed via the disk accretion \citep{Reid2007, Testi2010, Plambeck2013}. 
Nevertheless, the estimated parameters of the disk, such as inner/outer radii of the ring-like structure, enclosed mass or rotationally supported mass, and excitation temperature, are still uncertain because of a lack of spatially resolved data. 
In fact, we still see some deviation between the model and observed results in PV diagram and spectral profile. 
Further higher angular resolution observations with ALMA of the multi-transition H$_{2}$O lines will be crucial to shed light on more detailed physical and dynamical properties of the massive protostar Source~I and its surrounding circumstellar disk. 

\acknowledgements
We are grateful to T. Hosokawa, M. Momose, H. Nomura, N. Sakai, and S. Yamamoto for valuable discussions and K. Hada and A. Kataoka for useful comments. 
This letter makes use of the following ALMA data: ADS/JAO.ALMA\#2011.0.00199.S. 
ALMA is a partnership of ESO (representing its member states), NSF (USA) and NINS (Japan), together with NRC (Canada) and NSC and ASIAA (Taiwan), in cooperation with the Republic of Chile. 
The Joint ALMA Observatory is operated by ESO, AUI/NRAO and NAOJ. 
T.H. is supported by the MEXT/JSPS KAKENHI Grant Numbers 21224002, 24684011, and 25108005, and the ALMA Japan Research Grant of NAOJ Chile Observatory, NAOJ-ALMA-0006. 
M.H. is supported by the MEXT/JSPS KAKENHI Grant Numbers 24540242 and 25120007. 

{\it Facilities:} \facility{ALMA}.

\end{document}